\documentclass[twocolumn,showpacs,preprintnumbers,amsmath,amssymb]{revtex4}
\usepackage[T1]{fontenc}
\usepackage[latin1]{inputenc}


\input{tcilatex}

\begin{document}

\title{Checking for optimal solutions in some $NP$-complete problems}
\author{Michel Bauer and Henri Orland} \affiliation{Service de
  Physique Théorique, CEA-Saclay, 91191 Gif-sur-Yvette cedex, France}

\date{\today}

\begin{abstract}
  For some weighted $NP$-complete problems, checking whether a
  proposed solution is optimal is a non-trivial task. Such is the case
  for the celebrated traveling salesman problem, or the spin-glass
  problem in 3 dimensions. In this letter, we consider the weighted
  tripartite matching problem, a well known $NP$-complete problem. We
  write mean-field finite temperature equations for this model, and
  show that they become exact at zero temperature. As a consequence,
  given a possible solution, we propose an algorithm which allows to
  check in a polynomial time if the solution is indeed optimal. This
  algorithm is generalized to a class of variants of the multiple
  traveling salesmen problem.
\end{abstract}


\maketitle

A combinatorial optimization problem is defined as the minimization of
a cost function over a discrete set of configurations \cite{Pap_Ste}.
Typically, in statistical physics, finding the ground state of an
Ising spin-glass (Ising model with random interactions) is a
combinatorial optimization problem, where the cost function is the
magnetic energy of the system. The size $N$ of an optimization problem
is the number of degrees of freedom of the system over which the
minimization is performed. In a spin-glass problem, the size $N$ is
just the number of spins. In the following, we shall be interested
only in problems for which the cost function can be calculated in a
time which is polynomial in the size $N$.  This very wide class of
problems is called the $NP$ class. Many physical problems belong to
this class.

An optimisation problem {\cal Q} is said to be $NP$-complete if it is in $NP$,
and if all problems in $NP$ can be shown to be polynomially algorithmically
reducible to {\cal Q}.
Therefore, $NP$-complete problems are in
some sense the most difficult $NP$ problems. All $NP$-complete
problems are algorithmically polynomially equivalent. 
The archetype of such
problems is the celebrated traveling salesman problem (TSP): Given $N$
cities, find the shortest path going through each city once and only
once. Although many algorithms exist which provide exact solutions for
small enough $N$ or almost optimal solutions for larger $N$, there is
no known algorithm which provides the exact shortest path in a
polynomial time.  This is due to the combinatorial complexity of the
paths, and to the strong non-convexity of the cost function in phase
space, which manifests itself by an exponentially large number of
local minima.

As stated above, any combinatorial optimization problem which is
algorithmically equivalent to the TSP is $NP$-complete. Famous
examples of such problems include the spin-glass (SG) problem in
dimension $d\geq3$ \cite{Bah}, the Hamiltonian path (HP) problem (given a graph,
find a path which goes through each points of the graph once and only
once), the weighted tripartite matching problem (TMP),
3-satisfiability, etc. For an extensive list of $NP$-complete
problems, see ref. \cite{Gar_Joh}.

In the spin-glass problem, physicists and computer scientists have
developed algorithms which allow to compute the exact ground state for
small enough systems \cite{Har}. The analysis of these exact ground
states and low-lying excited states allows to check the validity of
various spin-glass theories \cite{Har_Dom}.

If there is a polynomial time algorithm to solve a combinatorial
optimization problem, the problem is said to belong to the $P$-class.
Many problems belong to $P$, among which the assignment problem (also
known as the weighted bipartite matching problem denoted BMP), the
spin-glass problem in $d=2$, etc.  Of course the $P$ class is included
in the $NP$ class, but it is not known if the inclusion is
strict. This is the celebrated ``Is $P=NP$ ?'' problem.

The strong analogy between combinatorial optimization problems and the
physics of disordered systems was recognized in the early 1980s
\cite{Kir_Gel_Vec}, and was the basis of the development of simulated
annealing techniques in optimization. At finite temperatures, many of
these optimization problems exhibit glassy behaviour as seen in
disordered systems
\cite{Mez_Van,Orl,Mez_Par1,Mez_Par_Vir,Kir_Mon,Mar_Mez}.

As mentioned above, when going to larger sizes, finding exact ground
states of $NP$-complete problems becomes impossible, and one resorts
to non-deterministic methods like Monte Carlo algorithms or Markov
chains.  These methods provide candidates for solutions to the
optimization problem.  In some cases, checking that one has a true
solution of the optimization problem is an easy task. For instance, in
the HP problem, given a path, it is easy to check whether this path is
Hamiltonian and thus solves the problem. In many weighted problems
however, checking that one has a solution is a very non trivial task.
Such is the case for example in the TSP problem.  Given a tour, there
is no known algorithm to determine whether it is the optimal path,
except by actually computing the optimal path.
Similarly, given a set of coupling $J_{ij}$ and a spin configuration $%
S_{i}$, there is no known algorithm to check that this configuration
is the ground state of the spin-glass, except by computing its actual
ground state.

In the present paper, we study the weighted multipartite matching
problem.  To simplify, we specialize to the tripartite case. Before
defining it, we first recall the assignment problem or BMP (also
called the wedding problem). Assume we have two sets $\{ i=1,...,N\}$
and $\{ j=1,...,N\}$. A positive pairing cost $l_{ij}$ is assigned to
each matching of $i$ and $j$.  Note that the matrix $l_{ij}$ has no
reason to be taken symmetric. A matching of the two sets is in fact a
permutation $P$ of the set $\{ j\}$ and the corresponding cost is

\begin{equation}
L=\sum_{i=1}^{N}l_{iP(i)}  \label{1}
\end{equation}

Solving the BMP amounts to finding the permutation $P$ which minimizes
(\ref{1}), i.e. the complete pairing of $\{ i\}$ and $\{ j\}$ which
has lowest cost.

The simplest case of multipartite matching problem is the tripartite
one.  The TMP involves three sets $\{ i=1,...,N\}$ , $\{ j=1,...,N\}$
, $\{ k=1,...,N\}$ and a positive cost function $l_{ijk}$ . A matching
of the 3 sets is defined by a permutation $P$ of the set $\{ j\}$ and
a permutation $Q$ of the set $\{ k\}$ with cost

\begin{equation}
L=\sum_{i=1}^{N}l_{iP(i)Q(i)}  \label{2}
\end{equation}
Solving the TMP amounts to finding the permutations $P$ and $Q$ which
minimize (\ref{2}). Generalization to multipartite matchings is
straightforward.  As mentioned above, the bipartite version is
polynomial, whereas the tri-, quadri-, etc. matching are
$NP$-complete. Note that in the BMP, the number of possible
configurations is $N!$ whereas it is $(N!)^{2}$ in the TMP.  The phase
diagram of these multipartite matching problems has been studied
recently for sets of random independent costs \cite{Mar_Mez}.

In this article, we first formulate the finite temperature TMP as an
integral over complex fields. The saddle-point equations are derived
and are shown to provide the exact solution to the TMP at zero
temperature.  Unfortunately, this does not help in solving the
problem. However, from these equations, we find that given a
tripartite assignment, one can check whether it is the absolute
minimum of the cost function in a polynomial time. This method can
easily be applied to a whole class of multiple traveling salesmen
problems \cite{MTSP}.  This of course does not violate the
$NP$-complete character of these problems. We are not aware of any
other {\em weighted} $NP$-complete problems which could be checked in
a polynomial time, and hopefully this method could be generalized to
other problems.

It is possible to give an integral formulation of the TMP using
techniques similar to those used for the BMP \cite{Orl,Mez_Par1}.  We
want to calculate the partition function $Z$ of the TMP at temperature
$T$. Denoting by $k_{B}$ the Boltzmann constant and $\beta=1/k_{B}T$ ,
we have

\begin{equation}
Z=\sum_{P,Q\in S_{N}}\exp\left(-\beta\sum_{i=1}^{N}l_{iP(i)Q(i)}\right)
\label{part}
\end{equation}
where $S_{N}$ denotes the group of permutations of $N$ objects (symmetric
group). We define

\[
U_{ijk}=\exp\left(-\beta l_{ijk}\right)
\]

\begin{eqnarray}
Z&=&\int \prod_{i=1}^N {d\phi_{i}d\phi_{i}^* \over \pi}
{d\psi_{i} d\psi_{i}^* \over \pi}
{d\chi_{i} d\chi_{i}^* \over \pi}
\nonumber \\
&&\times \exp\left(-\sum\left({\phi_{i}}\phi_{i}^*+%
{\psi_{i}}\psi_{i}^*+{\chi_{i}}\chi_{i}^*\right)\right)  \nonumber \\
&&\times
\exp\left(\sum_{i,j,k=1}^{N}U_{ijk}\phi_{i}\psi_{j}\chi_{k}\right)
\prod_{i=1}^{N}\left({\phi_{i}^*}{\psi}_{i}^*
{\chi_{i}^*}\right) 
\label{real}
\end{eqnarray}

Let us show that this is an exact expression for the partition
function of the TMP. We use Wick's theorem. We first note that the
contraction of each field with its conjugate is just equal to 1.
Expanding the exponential in powers of $U_{ijk}$, we should contract
each $\phi_i, \psi_i, \chi_i$ with its conjugate $\phi_i^*, \psi_i^*,
\chi_i^*$. Since the integrand is linear in each $\phi_i^*, \psi_i^*,
\chi_i^*$, the expansion in powers of $U_{ijk}$ should be limited to
first order. This obviously proves equation (\ref{real}).

A similar formula can easily be obtained for the BMP, involving only
two kinds of fields instead of three.

A standard way to approximate eq. (\ref{real}) is to perform a saddle-point
expansion. The saddle-point equations read

\[
\frac{1}{\phi_{i}^*}=\phi_{i} 
\]

\[
\phi_{i}^*=\sum_{j,k=1}^{N}U_{ijk}\psi_{j}\chi_{k} 
\]
and similar equations for the other variables. Eliminating the
conjugate variables, we have the $3N$ equations

\begin{equation}
1=\phi_{i}\sum_{j,k=1}^{N}U_{ijk}\psi_{j}\chi_{k}  \label{MF}
\end{equation}
for any $i$ and the analogous equations for the other variables.

There are $3N$ saddle-point equations for the $3N$ variables $%
\phi_{i},\psi_{i},\chi_{i}$. These equations are complicated
non-linear equations and can be solved numerically.  They differ from
those obtained from the cavity method \cite{Mez_Par2,Mar_Mez}.
However, they display some interesting properties in the zero
temperature limit. Introducing new variables

\[
\phi_{i}=e^{\beta a_{i}},\psi_{i}=e^{\beta b_{i}},\chi_{i}=e^{\beta c_{i}} 
\]
one can see that solving equations (\ref{MF}) in the zero temperature
limit amounts to finding two permutations $P$ and $Q$ such that

\begin{equation}
a_{i}+b_{P(i)}+c_{Q(i)}=l_{iP(i)Q(i)}  \label{o1}
\end{equation}
for any $i$ , and

\begin{equation}
a_{i}+b_{j}+c_{k} <  l_{ijk}  \label{o2}
\end{equation}
for any triplet $(i,j,k) \ne (i,P(i),Q(i))$ . The total cost of the
matching is

\begin{equation}
L=\sum_{i=1}^{N}l_{iP(i)Q(i)}=\sum_{i=1}^{N}\left( a_{i}+b_{i}+c_{i}\right) 
\end{equation}

We now proceed to prove that these equations are exact, i.e. their
solutions (if they exist) provide the optimal tripartite matching. We
first note that if we have two sets of solutions $(a,b,c)$ and
$(a',b',c')$ which satisfy equations (\ref{o1}) and (\ref{o2}) with
the same $P$ and $Q$, they necessarily have the same total cost, due
to equations (\ref{o1}).  Now consider another pair of permutations
$P^{\prime}$ and $Q^{\prime}$.  The total cost $L'$ associated to
these permutations is
\[
L^{\prime }=\sum_{i=1}^{N}l_{iP^{\prime }(i)Q^{\prime
}(i)}
\]

According to eq. (\ref{o2})
\[
a_i+ b_{P'(i)}+c_{Q'(i)} \le l_{iP^{\prime }(i)Q^{\prime }(i)}
\]
and therefore, summing over $i$ implies
\[
L \le L'
\]

Therefore, any matching other than $(P,Q)$ have larger cost,
which proves that $(P,Q)$ generates the optimal matching.

A formulation similar to (\ref{o1}) and (\ref{o2}) has been known for
the BMP, with only two sets of variables $a$ and $b$. It can be shown
that these equations can be solved using the so-called Hungarian
method \cite{Hun} which is an $O(N^{3})$ algorithm.

In the case of the TMP, these equations unfortunately cannot be solved
in a polynomial time. However, as we shall now see, they allow to
check in a polynomial time whether a proposed solution is the actual
optimal matching.

Let us consider the reciprocal problem: Assume we are given a matching. How
can we check whether it is optimal?

The matching is defined by a set of $N$ costs $\{l_{iP(i)Q(i)}\}$. 

If this set is optimal, then there exists a set of $3N$ variables
$a_{i},b_{i},c_{i}$ such that the constraints (\ref{o1}) and
(\ref{o2}) would be satisfied.

These are a set of $N$ equations and $N^{3}-N$ linear inequalities for
$3N$ variables. We can simplify further by using equations (\ref{o1}),
and obtain a set of $N^{3}-N$ inequalities for the $2N$ variables
$a_{i}$ and $b_{i}$

\begin{equation}
a_{i}+b_{j}-a_{Q^{-1}(k)}-b_{PQ^{-1}(k)}\leq l_{ijk}-l_{Q^{-1}(k)PQ^{-1}(k)k}  \label{ineq}
\end{equation}

This set of linear inequalities with integer coefficients belongs to
the well known class of optimization problems called "linear
inequalities" \cite{Pap_Ste}.  This problem, which is related to
linear programming, is known to be solvable in a time which is
polynomial in the size $2N$ of the problem \cite{ellipsoid,interior}.
Therefore, we can find, in a polynomial time, i) either that there is
a solution to (\ref{ineq}), in which case the proposed solution is the
optimal solution to the TMP, ii) or that there is no solution to these
inequalities, in which case the proposed solution is not the optimal
one.

We have not found a general proof of existence of the $a_i,b_i,c_i$
for the optimal solution.

Let us show how this method can be generalized to some variants of the
multiple traveling salesmen problem (MTSP). The MTSP is similar to the
TSP, except that the number of travelers in not equal to one.
Consider a set of $N$ points (in an abstract space) with positive
distances $l_{ij}$. We consider the TMP in which
\[
l_{i,j,k} = \frac{1}{2} (l_{ij}+l_{jk})
\] 
if $(i,j,k)$ are distinct
\[
l_{i,j,k} = \infty 
\]
if 2 indices are equal.

With this choice, a finite cost tripartite assignment can be viewed as
a set of loops visiting each point once and only once.  The
corresponding TMP thus amounts to a MTSP, with any number of salesmen,
each visiting at least 3 cities. If we are presented with a possible
optimal path, say $\{l_{12},l_{23},...,l_{N1}\}$, we look for two sets
of variables $a_i$ and $c_i$ which satisfy the linear inequalities:
\[
a_{i}+c_{k}-a_{j-1}-c_{j+1}\leq \frac{1}{2} (l_{ij}+l_{jk}-l_{j-1,j}-l_{j,j+1})
\]

Again, although the problem is $NP$-complete, checking the optimality
of the solution can be achieved in a polynomial time. The whole method
can be easily generalized to MTSP with loop sizes greater than $p$ in
terms of $p$-partite weighted matching problems, to find criteria of
optimality which are polynomial.

We have proposed an integral representation of the TMP, from which we
derive some mean-field equations.  These equations turn out to be
exact at zero temperature, and could have in fact been derived
directly without going through the mean-field method. However, we find
this approach interesting, as it might be generalized to other
$NP$-complete problems. The zero temperature equations cannot be
solved in a polynomial time.  However, given a test solution to the
problem, they allow to check in a polynomial time whether the proposed
solution is indeed the optimum of problem. This situation is quite
unique in $NP$-complete problems, since in principle, for such
problems, the existence of an exponentially large number of local
minima prevents checking the optimality in a polynomial time.  A
generalization of this method to the TSP or to the spin-glass problem
would be of great interest for the physics of disordered systems and
is currently under investigation.


\begin{thebibliography}{99}
\bibitem{MTSP} The multiple traveling salesmen problem is the same as
  the TSP, except that several loops (or several salesmen) are allowed,
  whereas only one is allowed in the TSP.
\end{thebibliography}
\end{document}